\documentclass[12pt]{article}

\usepackage{amsmath}
\usepackage{amssymb}
\usepackage{latexsym}
\usepackage{graphicx}
\usepackage{epsfig}

\def\beq{\begin{equation}}
\def\eeq{\end{equation}}
\def\bea{\begin{eqnarray}}
\def\eea{\end{eqnarray}}

\begin{document}

\begin{titlepage}

\vspace*{1cm}
\begin{center}
{\bf \Large Schwarzschild-de Sitter spacetime: the role of}
\vspace*{0.1cm}

{\bf \Large Temperature in the emission of Hawking radiation}

\bigskip \bigskip \medskip

{\bf Thomas Pappas}\footnote{Email: thpap@cc.uoi.gr}
and {\bf Panagiota Kanti}\footnote{Email: pkanti@cc.uoi.gr}

\bigskip
{\it Division of Theoretical Physics, Department of Physics,\\
University of Ioannina, Ioannina GR-45110, Greece}

\bigskip \medskip
{\bf Abstract}
\end{center}
We consider a Schwarzschild-de Sitter (SdS) black hole, and focus on the emission
of massless scalar fields either minimally or non-minimally coupled to gravity.
We use six different temperatures, two black-hole and four effective ones for
the SdS spacetime, as 
the question of the proper temperature for such a background is still debated in
the literature. We study their profiles under the variation of the cosmological
constant, and derive the corresponding Hawking radiation spectra. We demonstrate
that only few of these temperatures may support significant emission of radiation.
We finally compute the total emissivities for each temperature, and show that
the non-minimal coupling constant of the scalar field to gravity also affects
the relative magnitudes of the energy emission rates.

\end{titlepage}

\setcounter{page}{1}

\section{Introduction}
\label{sec:intro}

The emission of Hawking radiation \cite{Hawking} from black holes has been a favourite
topic in the literature since it combines the most fascinating objects of the General
Theory of Relativity with the manifestation of a quantum effect in curved spacetime. 
One of the oldest black-hole solutions is the Schwarzschild-de Sitter (SdS) solution
\cite{Tangherlini} describing a static, spherically-symmetric, uncharged black hole
formed in the presence of a positive cosmological constant. The literature on the
emission of Hawking radiation from such a black hole has been scarce: this topic was
first discussed in the seminal work by Gibbons and Hawking \cite{GH1}, while an early
attempt to calculate the particle production rate in SdS spacetime
appeared in \cite{KT}. A subsequent work studied the interaction of the Hawking radiation
emitted by a SdS black hole with a static source \cite{Castineiras}.

Remarkably, most of the works focusing
on the emission of Hawking radiation from such a black hole have dealt with the
higher-dimensional version of the SdS spacetime.  The first such works studied
the emission of either scalars \cite{KGB, Labbe} or fields with arbitrary spin
\cite{Wu} both on the brane and in the bulk, while another work \cite{Harmark} 
studied analytically the scalar greybody factor in an arbitrary number of
dimensions. The emission of Hawking radiation from a purely 4-dimensional
SdS black hole, in the form of scalar fields either minimally or non-minimally
coupled to gravity, was studied only a few years ago \cite{Crispino}, and soon
afterwards the same study was extended in a higher-dimensional context 
\cite{KPP1, KPP3}. In a few additional works \cite{Anderson, Sporea, Ahmed,
Fernando, Boonserm}, the greybody factors for fields propagating in variants
of a Schwarzschild-de Sitter background was also studied.

However, the thermodynamics of the SdS spacetime has not been free of problems
and open questions. The temperature of the black-hole horizon, or otherwise the {\it bare}
temperature defined in terms of the corresponding surface gravity \cite{GH1}, fails
to take into account the absence of an asymptotically-flat limit. A {\it normalised}
temperature proposed in \cite{BH} resolves this problem. However, a new one soon
emerged: the SdS spacetime possesses a second horizon, the cosmological horizon,
that has its own temperature \cite{GH2}. An observer located at a point between
the two horizons will constantly interact with both of them, and thus will never be in
a true thermodynamic equilibrium. This problem may be ignored in the limit of a small
cosmological constant, when the two horizons are located far away, but it worsens
when the cosmological constant takes a large value. In a new approach adopted
in a series of works \cite{Shankar, Urano, Lahiri, Bhatta, LiMa} (see also \cite{Mann_review}
for a review), the notion of the {\it effective temperature} for the SdS spacetime 
was proposed that implements both the black-hole and the cosmological horizon
temperatures (for a number of additional works on SdS thermodynamics, see
\cite{Romans}-\cite{Pourhassan}). 

For the study of the emission of Hawking radiation by a SdS black hole, it was either the
normalised \cite{KGB, Wu, KPP3} or bare \cite{Labbe, Crispino} temperatures that were
used. Although the claim was made (see, for example, \cite{Urano}) that an
effective temperature should be used instead, until recently no such work existed in
the literature. In \cite{KTP1}, we undertook this task and performed a comprehensive
study of the radiation spectra for a higher-dimensional SdS black hole, by using five
different temperatures. We demonstrated that the energy emission rates depend strongly
on the choice of the temperature as also does the corresponding bulk-over-brane 
energy ratio. Here, we perform a similar study for the case of a purely 4-dimensional
SdS spacetime for the following two reasons: (i) the first effective temperature that
appeared in the literature was formulated for a 3-dimensional space and thus may be
implemented only in the context of a 4-dimensional analysis, (ii) our previous study
\cite{KTP1} showed that some of the temperatures considered tend to acquire similar
profiles as the number of spacelike dimensions increases; thus, we expect the largest
differences to appear for the lowest dimensionality, i.e. for $D=4$.

The outline of our paper is as follows: in Section 2, we present
the gravitational background, and perform a detailed study of the different
definitions of the temperature of a Schwarzschild-de Sitter spacetime. In
Section 3, we consider a field theory of a scalar field that may be either
minimally or non-minimally coupled to gravity, and solve its equation of
motion in the aforementioned background in a exact numerical way. Its greybody
factor is then used to derive the energy emission rates from the SdS
black hole, for both a minimal and non-minimal coupling to gravity, and for
the different temperatures. In Section 4, we calculate and compare the total
emissivities of the black hole in each case and, in Section 5, we summarise
our conclusions.


\section{The Gravitational Background}
\label{subsec:framework}

We consider the Einstein-Hilbert action in four dimensions and assume
also the presence of a positive cosmological constant $\Lambda$. Then,
the gravitational action reads 
\beq
S_G=\int d^{4}x\, \sqrt{-g}\,\left(\frac{R}{2 \kappa^2} - \Lambda\right)\,,
\label{action}
\eeq
where $R$ is the Ricci scalar, $\kappa^2=8\pi G$, and $g$ the determinant
of the metric tensor $g_{\mu\nu}$. By varying the above action with respect
to $g_{\mu\nu}$, we obtain the Einstein's field equations that have the
following form:
\beq
R_{\mu\nu}-\frac{1}{2}\,g_{\mu\nu}\,R = -\kappa^2 g_{\mu\nu} \Lambda\,.
\label{field_eqs}
\eeq
It is well-known that the above equations admit a spherically-symmetric
solution of the form \cite{Tangherlini}
\beq
ds^2 = - h(r)\,dt^2 + \frac{dr^2}{h(r)} + r^2\,(d\theta^2 +
\sin^2\theta\,d\varphi^2)\,,
\label{bhmetric}
\eeq
where the radial function $h(r)$ is given by the expression
\beq
h(r) = 1-\frac{2M}{r} - \frac{\kappa^2 \Lambda}{3}\,r^2\,.
\label{h-fun}
\eeq
The above solution describes a Schwarzschild-de-Sitter (SdS) spacetime, with the
parameter $M$ being the black-hole mass. The horizons of the SdS spacetime follow
from the equation $h(r)=0$, which yields two real, positive roots
for $0 < \Lambda M^2/9 <1$ \cite{GH1, Molina}. The smaller of the two roots stands
for the black-hole horizon $r_h$, and the larger for the cosmological horizon $r_c$.
In the critical limit $\Lambda M^2/9=1$, known also as the Nariai limit \cite{Nariai},
the two horizons coincide and are given by $r_h=1/\sqrt{\Lambda}=r_c$.

In principle, the temperature of a black hole is defined in terms of
its surface gravity $k_h$ at the location of the horizon \cite{GH1, GH2} given by
the covariant expression
\beq
k_h^2 =-\frac{1}{2}\,\lim_{r \rightarrow r_h} (D_\mu K_\nu)(D^\mu K^\nu)\,,
\label{surface_grav1}
\eeq
where $D_\mu$ is the covariant derivative and 
$K=\gamma_t\,\frac{\partial \,\,}{\partial t}$ the timelike Killing vector
with $\gamma_t$ a normalization constant. For a spherically-symmetric
gravitational background, $k_h$ is simplified to \cite{York}
\beq
k_h=\frac{1}{2}\,\frac{1}{\sqrt{-g_{tt} g_{rr}}}\,|g_{tt,r}|_{r=r_h}\,,
\label{surface_grav2}
\eeq
and the temperature of the Schwarzschild-de Sitter black hole (\ref{bhmetric})
finally takes the form \cite{GH1, York}
\beq
T_0 =  \frac{k_h}{2\pi}=\frac{1-\Lambda r_h^2}{4\pi r_h}\,.
\label{Temp0}
\eeq
In the above, we have used the condition $h(r_h)=0$ to replace $M$ in terms of
$r_h$ and $\Lambda$, and set $\kappa^2=1$ for simplicity.

However, the SdS spacetime (\ref{bhmetric}) does not have an asymptotic flat limit,
where traditionally all parameters of a black hole are defined: the metric function
$h(r)$ interpolates between two zeros, at $r_h$ and $r_c$, reaching a maximum
value at an intermediate point $r_0$ given by the expression $r_0^3=3M/\Lambda$;
there, $h(r_0)=1-\Lambda r_0^2$, that indeed deviates from unity the larger $\Lambda$
is. In order to fix this problem, a `normalised' expression for the temperature of a
SdS black hole was proposed in \cite{BH}, given by
\beq
T_{BH} = \frac{1}{\sqrt{h(r_0)}}\,\frac{1-\Lambda r_h^2}{4\pi r_h}\,.
\label{TempBH}
\eeq
Mathematically, the inclusion of the factor $\sqrt{h(r_0)}$ is dictated by the
non-trivial normalisation constant $\gamma_t$ in the expression of the Killing
vector $K^\mu$ when the latter is defined away from a flat spacetime. From the physical
point of view, it is at the point $r_0$ that the effects of the black-hole
and cosmological horizons cancel out and thus the point the closest to an
asymptotically flat limit.

Nevertheless, the thermodynamics of a SdS black hole faces another problem: 
one may define the surface gravity $k_c$ of the cosmological horizon in a 
similar way and, from that, the corresponding temperature \cite{GH1, GH2}
\beq
T_c = - \frac{k_c}{2\pi}=-\frac{1-\Lambda r_c^2}{4\pi r_c}\,.
\label{Tempc}
\eeq
For small values of the cosmological constant, the two horizons are located
far way from each other, and one may develop two independent thermodynamics
\cite{BH, GH2, Sekiwa}. But, as $\Lambda$ increases while keeping $M$ fixed,
the two horizons approach each other finally becoming coincident at the critical
limit; as the two temperatures, $T_0$ and $T_c$, are in principle different,
an observer located at an arbitrary point of the causal region $r_h<r<r_c$ 
interacting with both horizons will never be in a true thermodynamical
equilibrium. 

As a result, the concept of the {\it effective temperature} of the Schwarzschild-de Sitter
spacetime, that involves both temperatures $T_0$ and $T_c$, has emerged during the
recent years.  In the first approach that was taken in the literature \cite{Urano}, 
a thermodynamical first law for a Schwarzschild-de Sitter spacetime was written
by applying the extended Iyer-Wald formalism \cite{Wald}: in this, it was assumed that
the black-hole mass plays the role of the internal energy of the system ($M=E$),
the entropy is the sum of the entropies of the two horizons ($S=S_h+S_c$) and
the volume is the one of the observable part of spacetime ($V=V_c-V_h$). Then,
the coefficient of $\delta S$ in the first law was identified with the effective temperature
of the system and found to be:
\beq
T_{effEIW}=\frac{r_h^4\,T_c+r_c^4\,T_0}{(r_h+r_c)\,(r_c^3-r_h^3)}\,.
\label{Teff-EIW}
\eeq
In the second approach taken \cite{Shankar, Lahiri, Bhatta, LiMa} (for a nice review on both
approaches, see \cite{ Mann_review}), it was assumed instead that the black-hole mass
plays the role of the enthalpy of the system ($M=-H$), the cosmological constant that of
the pressure ($P=\Lambda/8\pi$) while the entropy is still $S=S_h+S_c$. In that case,
the effective temperature of the system was found to have the expression
\beq
T_{eff-}=\left(\frac{1}{T_c}-\frac{1}{T_0}\right)^{-1}=\frac{T_0 T_c}{T_0-T_c}=
-\frac{(1 -\Lambda r_h^2)\,(1-\Lambda r_c^2)}
{4\pi\,(r_h +r_c)\,(1-\Lambda r_h r_c)}\,.
\label{Teff-}
\eeq

However, the latter effective temperature (\ref{Teff-}) is not always positive-definite and
may exhibit infinite jumps near the critical point in charged versions of the SdS spacetime
\cite{Mann_review}. An alternative expression for the effective temperature of the SdS
spacetime was thus proposed in \cite{Mann_review} (see also \cite{Shankar}) of the form
\beq
T_{eff+}=\left(\frac{1}{T_c}+\frac{1}{T_0}\right)^{-1}=\frac{T_0 T_c}{T_0 + T_c}
=-\frac{(1 -\Lambda r_h^2)\,(1-\Lambda r_c^2)}
{4\pi\,(r_c -r_h)\,(1+\Lambda r_h r_c)}\,.
\label{Teff+}
\eeq
The above expression could follow from an analysis similar to that leading to $T_{eff-}$ 
but assuming that the entropy of the system is the difference of the entropies of the two
horizons, i.e. $S=S_c-S_h$; however, no physical reason exists for that. In a recent work
of ours \cite{KTP1}, we observed instead that, if one followed the same approach that led
to $T_{eff-}$ but merely replaced the `bare' temperature $T_0$ with the `normalised' one
$T_{BH}$, one would obtain the following expression for the effective temperature of the
SdS spacetime
\beq
T_{effBH}=\left(\frac{1}{T_c}-\frac{1}{T_{BH}}\right)^{-1}=\frac{T_{BH} T_c}{T_{BH} - T_c}=
-\frac{(1 -\Lambda r_h^2)\,(1-\Lambda r_c^2)}
{4\pi\,(r_h\,\sqrt{h(r_0)} + r_c)\,(1- \Lambda r_h r_c)}\,.
\label{TeffBH}
\eeq
As we will shortly see, the above expression shares several characteristics with the $T_{eff+}$;
at the same time, it retains the usual assumption for the entropy of the system ($S=S_h+S_c$)
and takes into account the absence of asymptotic flatness.

\begin{figure}[t]
  \begin{center}
\includegraphics[width = 0.50 \textwidth] {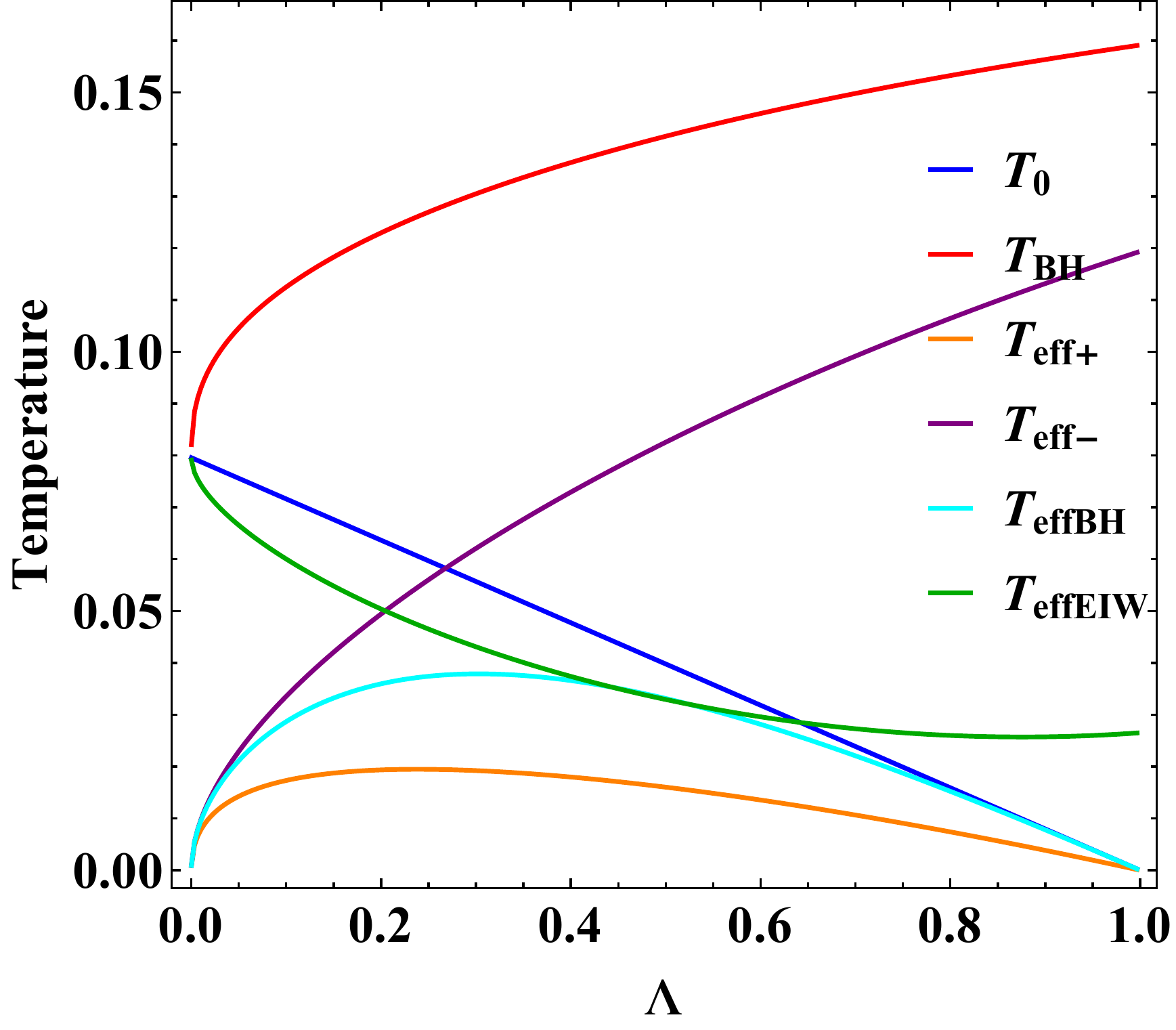}
    \caption{Temperatures for a Schwarzschild-de Sitter black hole (from top to bottom
in the low-$\Lambda$ regime: $T_{BH}$, $T_0$,  $T_{effEIW}$, $T_{eff-}$, $T_{effBH}$, and
$T_{eff+}$) as a function of the co- smological constant $\Lambda$.}
   \label{Temp_plot}
  \end{center}
\end{figure}

In Fig. \ref{Temp_plot}, we depict the behaviour of all six temperatures ($T_0$, $T_{BH}$,
$T_{effEIW}$, $T_{eff-}$, $T_{eff+}$, $T_{effBH}$) as functions of the cosmological constant
$\Lambda$. Throughout the present analysis, we will allow $\Lambda$ to vary in the
complete regime $[0, \Lambda_c]$, where $\Lambda_c=1/r_h^2$ is the maximum, critical
value of the cosmological constant for which the two horizons coincide - for simplicity, we
will keep the black-hole horizon fixed ($r_h=1$), and thus allow $\Lambda$ to take values
in the range $[0,1]$. Starting from the low-$\Lambda$ regime, we observe that the six
temperatures are split into two distinct groups that adopt two different asymptotic values
as $\Lambda \rightarrow 0$. The first group is comprised by the two black-hole
temperatures, $T_0$ and $T_{BH}$, and the effective temperature $T_{effEIW}$: the
first two temperatures naturally reduce to the temperature of the Schwarzschild black
hole $T_H=1/4\pi r_h$ when the cosmological constant vanishes; as one may see from
Eq. (\ref{Teff-EIW}), $T_{effEIW}$ also has the same smooth limit as $r_c \rightarrow \infty$.
We may thus conclude that $T_{effEIW}$ has been built on the assumption that the black-hole
horizon should always be present while $r_c$ may be located at either a finite or infinite
distance. On the contrary, the three effective temperatures  $T_{eff-}$, $T_{eff+}$, and
$T_{effBH}$ vanish when $\Lambda$ goes to zero (but reduce to $T_c$ when $r_h \rightarrow
0$); this is due to the fact that these effective temperatures were derived under the
assumption that $\Lambda$ is non-zero, standing for the pressure of the system -- 
in this approach, it is the cosmological horizon that should always be present by
construction whereas $r_h$ may vanish or not.

As the cosmological constant increases, the normalised black-hole $T_{BH}$ and
the effective temperature $T_{eff-}$ monotonically increase -- as shown in Fig.
\ref{Temp_plot} -- while the bare black-hole
one $T_0$ together with the effective temperature $T_{effEIW}$ monotonically decrease.
On the other hand, both effective temperatures $T_{eff+}$ and $T_{effBH}$ first increase
with $\Lambda$ and, after reaching a maximum value, start decreasing thus exhibiting
a similar behaviour. 

Towards the critical point ($\Lambda \rightarrow 1$), the six temperatures again
split into two groups: the first one is now comprised of the temperatures
$T_{BH}$, $T_{eff-}$ and $T_{effEIW}$ that retain an asymptotic, non-zero value
at the critical limit. We may easily justify this behaviour by looking at
Eqs. (\ref{TempBH}), (\ref{Teff-}) and (\ref{Teff-EIW}): in the limit $r_h \rightarrow 
r_c \rightarrow 1/\sqrt{\Lambda}$, both the numerator and denominator in all
three expressions go to zero in such a way that their ratio remains constant.
On the other hand, the remaining three temperatures, $T_0$, $T_{eff+}$, and
$T_{effBH}$, all vanish at the critical limit -- indeed, it is now only the numerators
in Eqs. (\ref{Temp0}), (\ref{Teff+}) and (\ref{TeffBH}) that go to zero while all
denominators are non-vanishing. 

Comparing their overall behaviour, one may immediately see the dominance of
the normalised temperature $T_{BH}$ over the whole $\Lambda$-regime. For
small cosmological constant, the values of the bare $T_0$ and effective
temperature $T_{effEIW}$ are also comparable. As the critical point
is approached, the dominance of $T_{BH}$ still holds, however, the effective
$T_{eff-}$ is also taking on a large value. Since the greybody factors for 
a specific type of field will be common, the aforementioned behaviour
will also determine the corresponding spectra of Hawking radiation when
each one of the aforementioned temperatures is used. In the next section,
we will take up this task and consider a scalar field, minimally- or
non-minimally coupled to gravity. We will study its emission in a 4-dimensional
SdS spacetime, and determine the radiation spectra -- and their profiles in
terms of $\Lambda$ -- by using each one of the above temperatures.


\section{The Effect of the Temperature on the EERs}
\label{sec:plots}

In the recent work of ours \cite{KTP1}, where the radiation spectra for a higher-dimensional
SdS black hole were studied in detail using five different temperatures, it became clear
that the temperature profiles depend strongly on the dimensionality of spacetime. In fact,
the differences between the two black-hole temperatures, $T_0$ and $T_{BH}$, and 
between the three effective temperatures, $T_{eff-}$, $T_{eff+}$ and $T_{effBH}$,
were amplified as $n$ decreased. Thus, here, we turn to the calculation of the energy
emission rates (EERs) for the case of a purely 4-dimensional SdS black hole.
We will also include a fourth effective temperature, $T_{effEIW}$, that was left out
in the analysis of \cite{KTP1}: the reason for that was that its expression (\ref{Teff-EIW})
explicitly involves the 3-dimensional spatial volume, and thus its generalisation in a
higher-dimensional spacetime needs a careful consideration.

Let us consider the following field theory describing a massless, scalar field with
a non-minimal coupling to gravity
\beq
S_\Phi=-\frac{1}{2}\,\int d^{4}x \,\sqrt{-g}\left[\xi \Phi^2 R
+\partial_\mu \Phi\,\partial^\mu \Phi \right]\,.
\label{action-scalar}
\eeq
In the above, $g_{\mu\nu}$ is the metric tensor defined in Eq. (\ref{bhmetric}), and $R$ the
scalar curvature $R=4\kappa^2 \Lambda$. Also, $\xi$ is a constant, with the value $\xi=0$
corresponding to the minimal coupling and the value $\xi=1/6$ to the conformal coupling. 
The equation of motion of the scalar field has the form
\beq
\frac{1}{\sqrt{-g}}\,\partial_\mu\left(\sqrt{-g}\,g^{\mu\nu}\partial_\nu \Phi\right)
=\xi R\,\Phi\,.
\label{field-eq}
\eeq
If we assume a factorized ansatz for the field, i.e. 
$\Phi(t,r,\theta,\varphi)= e^{-i\omega t}\,P(r)\,Y(\theta,\varphi)$,
where $Y(\theta,\varphi)$ are the scalar spherical harmonics, we obtain a
radial equation for the function $P(r)$ of the form
\beq
\frac{1}{r^{2}}\,\frac{d \,}{dr} \biggl(hr^{2}\,\frac{d P}{dr}\,\biggr) +
\biggl[\frac{\omega^2}{h} -\frac{l(l+1)}{r^2}-4 \xi \kappa^2 \Lambda\biggr] P=0\,.
\label{radial-xi}
\eeq
We note that the non-mininal coupling term acts as an effective mass term for the
scalar field \cite{Crispino, KPP1, KPP3}: any increase in its value increases the `mass'
of the field and thus suppresses both the greybody factors and radiation spectra
especially in the low-energy regime \cite{Page, Jung, Sampaio, KNP1}. The mass term
also depends on the value of $\Lambda$: in the minimal coupling case, $\Lambda$
enhances the EERs, however, for $\xi \neq 0$, the role of $\Lambda$ is more subtle
to infer and an exact computation of the radiation spectra is therefore necessary.

Equation (\ref{radial-xi}) was first studied in \cite{Crispino} and later extended in a
higher-dimensional context in \cite{KPP1, KPP3}. Here, we follow the analysis of the
last work, and focus on the 4-dimensional case with $n=0$. The analytic study of
Eq. (\ref{radial-xi}) in the near-horizon regime leads to a general solution written in
terms of a hypergeometric function. When expanded in the limit $r \rightarrow r_h$,
the solution takes the form of an ingoing free wave, namely
\beq
R_{BH} \simeq A_1\,f^{\alpha_1} = A_1\,e^{-i(\omega r_h/A_h)\,\ln f}\,,
\label{BH-exp}
\eeq
where $A_h=1-\Lambda r_h^2$,  and $f$ is a new radial variable defined by the
relation: $ r \rightarrow f(r) = h(r)/(1-\Lambda r^2/3)$. Upon setting the arbitrary
constant $A_1$ to unity, the above asymptotic solution leads to the conditions \cite{KPP3}
\beq
R_{BH}(r_h)=1\,, \qquad  \frac{dR_{BH}}{dr}\biggr|_{r_h} \simeq -\frac{i \omega}{h(r)}\,. 
\label{R_num}
\eeq
The above expressions serve as boundary conditions for the numerical integration of
Eq. (\ref{radial-xi}). 

The solution of Eq. (\ref{radial-xi}) near the cosmological horizon is again given in terms
of hypergeometric functions. Taking the limit $r \rightarrow r_c$, we now find \cite{KPP1, KPP3}
\beq 
R_C \simeq  B_1\,e^{-i (\omega r_c/A_c)\ln f} + 
B_2\,e^{i (\omega r_c/A_c)\ln f} \,, \label{CO-exp}
\eeq
where $A_c=1-\Lambda r_c^2$. The constant coefficients $B_{1,2}$ are easily identified with
the amplitudes of the ingoing and outgoing free waves. As a result, the greybody factor, or
transmission probability, for the scalar field may be expressed as
\beq
|A|^2=1-\left|\frac{B_2}{B_1}\right|^2\,.
\label{greybody}
\end{equation}
By numerically integrating Eq. (\ref{radial-xi}), we may find the $B_{1,2}$ coefficients:
we start the numerical integration close to the black-hole horizon,
i.e. from $r=r_h+\epsilon$, where $\epsilon=10^{-6}-10^{-4}$, and using the boundary
conditions (\ref{R_num}) we proceed towards the cosmological horizon. There, we isolate
the constant amplitudes $B_{1,2}$ (for more information on this, see \cite{KPP3}) and 
determine the greybody factor $|A|^2$. According to the results of our numerical
analysis, the greybody factor is suppressed with the non-minimal coupling constant 
$\xi$ over the whole energy regime; for small values of $\xi$, the cosmological constant
enhances the radiation spectra however, for large values of $\xi$, an increase in $\Lambda$
may cause a suppression especially in the low-energy regime.

We may now proceed to derive the differential energy emission rate for scalar fields from a
SdS black hole. The power emission spectrum is traditionally given by the expression
\cite{Hawking, KGB}
\beq
\frac{d^2E}{dt\,d\omega}=\frac{1}{2\pi}\,\sum_l\,\frac{N_l\,|A|^2\,\omega}{\exp(\omega/T)-1}\,,
\label{diff-rate-brane}
\eeq
where $\omega$ is the energy of the emitted particle, and $N_l=2l+1$ the multiplicity
of states that have the same angular-momentum number. The above formula describes a 
thermal spectrum that takes into account the back-scattering of the emitted modes, via
the presence of the greybody factor $|A|^2$. It has been used to describe the emission
of Hawking radiation from a plethora of four- and higher-dimensional black holes
(see \cite{Kanti:2004, KW} and references therein) as well as from a large number of
stringy or D-brane backgrounds (see, for example \cite{Das, Maldacena}). 

One should however be careful: although the authors of \cite{GH1} anticipated the
existence of a thermal spectrum of the form (\ref{diff-rate-brane}) for a SdS
background, the presence of the second (i.e. cosmological) horizon prevented them
from explicitly demonstrating that. The Schwarzschild and SdS spacetime have 
a number of similarities: they are both spherically-symmetric
and static. These two features allow us to define positive-frequency basis
modes -- that are necessary for the study of Hawking radiation -- in both backgrounds. 
In the Schwarzschild spacetime, the ``up'' modes \cite{KW, Birrell} are defined 
at the past event horizon while the ``in'' modes are defined at asymptotic infinity.
In the SdS spacetime, the asymptotically-flat regime is missing and replaced
by the cosmological horizon, and this seems to create a problem. But the calculation
of the particle production rate does not need a Minkowskian limit but only an
inertial observer \cite{Birrell}. And such an observer is always present in a
SdS spacetime residing at the point $r=r_0$: it is there that the effects of
the black-hole and cosmological horizons exactly cancel, and the proper
acceleration of the observer is zero. The only attempt in the literature to
calculate the particle production rate in a SdS spacetime \cite{KT} defined
the ``in'' modes close to the cosmological horizon, and found a non-thermal
spectrum -- that was a natural result, since a non-inertial observer fails to
detect a thermal spectrum \cite{Birrell, Ellis, Chakraborty}. 

It is worth noting that the point $r_0$ is present for all values of the
cosmological constant: when $\Lambda$ goes to zero, $r_0$ becomes the asymptotic
infinity; when $\Lambda \rightarrow \Lambda_c$, $r_0$ is approached on both
sides by $r_h$ and $r_c$ until they all match at the critical limit. 
Therefore, we expect an inertial observer residing at the point $r=r_0$ to
detect a Hawking radiation spectrum given indeed by Eq. (\ref{diff-rate-brane})
and for all values of the cosmological constant. In what follows, we will
use Eq. (\ref{diff-rate-brane}) where the temperature $T$ will be taken to
be equal, in turn, to $T_0$, $T_{BH}$, $T_{effEIW}$, $T_{eff-}$, $T_{eff+}$
and $T_{effBH}$, in order to derive the corresponding radiation spectra. 
The sum over the $l$-modes will be extended up to the $l = 7$
as all higher modes have negligible contributions to the total emission rate.

\begin{figure}[t!]
  \begin{center}
\mbox{\includegraphics[width = 0.49 \textwidth] {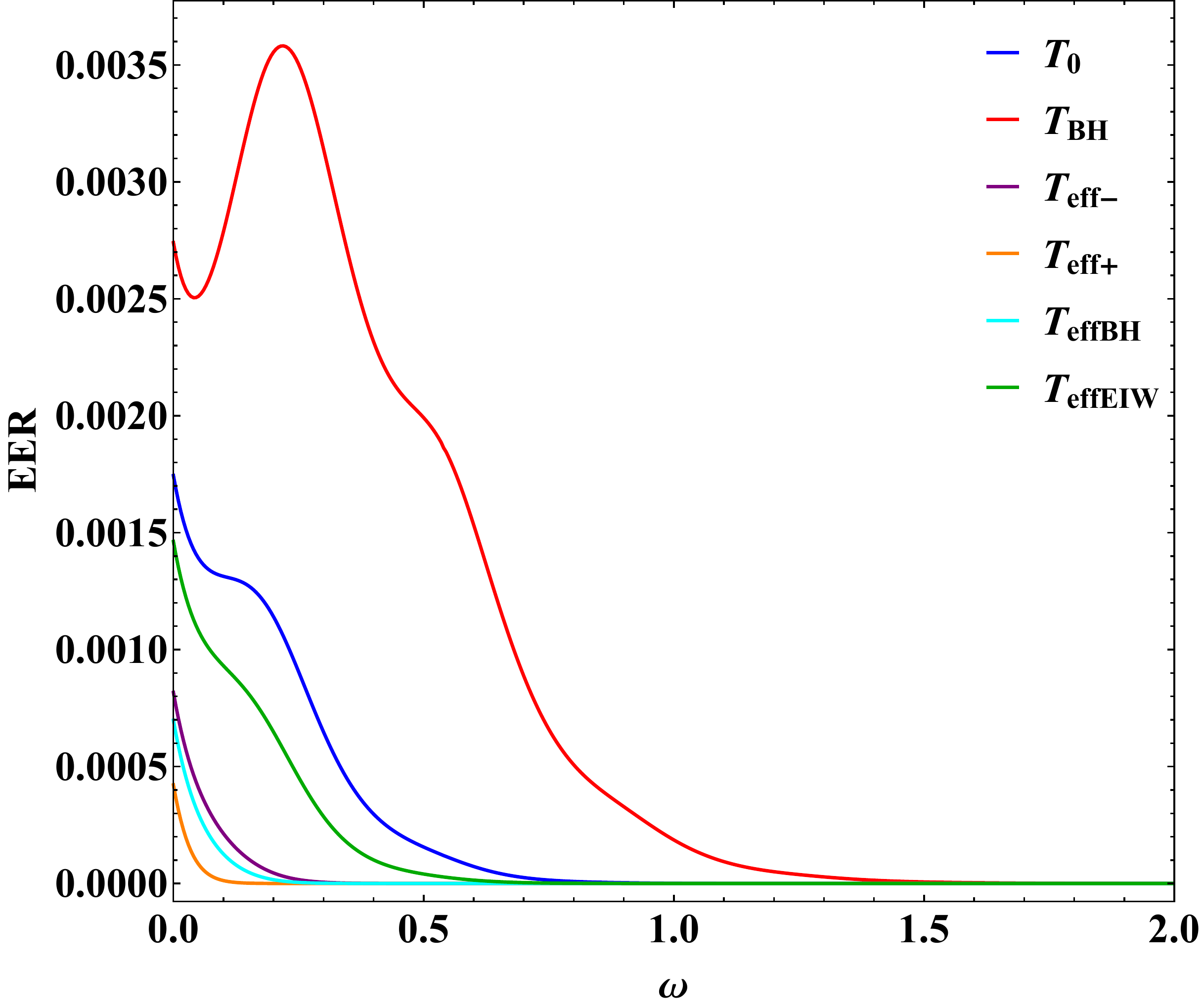}}
\hspace*{-0.1cm} {\includegraphics[width = 0.47 \textwidth]
{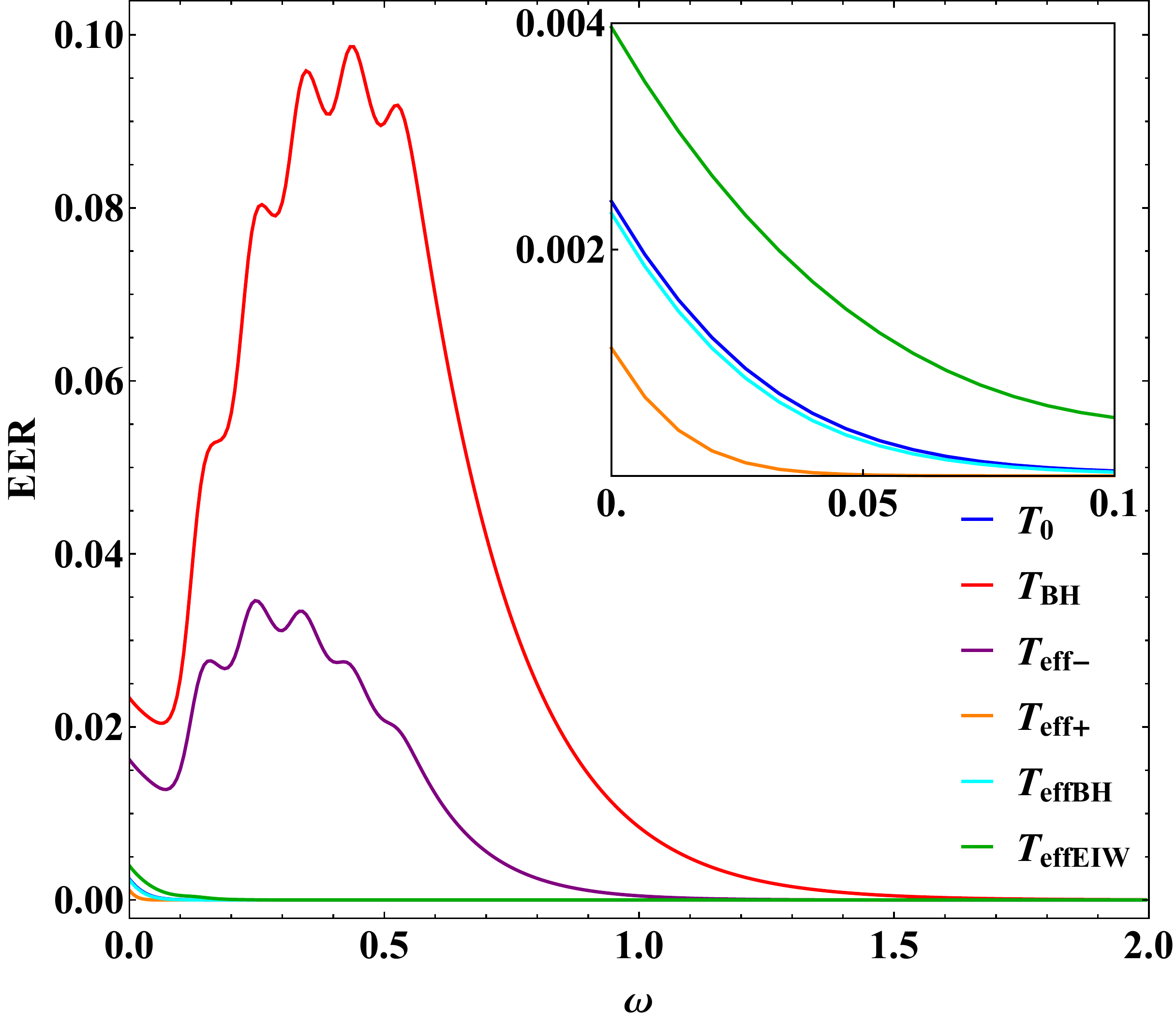}}
    \caption{Energy emission rates for minimally-coupled scalar fields from a
Schwarzschild-de Sitter black hole for:  \textbf{(a)} $\Lambda=0.1$ (in units of $r_h^{-2}$),
and  $T=T_{BH}$, $T_0$, $T_{effEIW}$, $T_{eff-}$, $T_{effBH}$, $T_{eff+}$ (from top to bottom),
and \textbf{(d)} $\Lambda=0.8$ and $T=T_{BH}, T_{eff-}$ (from top to bottom, again).}
   \label{EER_xi0}
  \end{center}
\end{figure}

Let us start with the minimal-coupling case, with $\xi=0$: the radiation spectra, for the
six temperatures and for two indicative values of the cosmological constant, are depicted
in Fig. \ref{EER_xi0}. Here, the effective mass term vanishes, and the emission curves
exhibit the characteristic feature of the non-vanishing asymptotic limit as $\omega 
\rightarrow 0$; this is due to the following non-vanishing geometric limit 
\beq
|A^2|=\frac{4 r_h^2r_c^2}{(r_c^2+r_h^2)^2}+ {\cal O}(\omega)\,,
\label{geom_brane}
\eeq
adopted by the greybody factor for a massless, free scalar field propagating in a 
SdS black-hole background \cite{KGB, Harmark, Crispino, KPP1, KPP3}. In fact, for the
low value $\Lambda=0.1$, the dominant emission channel lies in the low-energy regime. 
As $\Lambda$ increases, the energy emission curves are significantly enhanced and
reach their maximal points at intermediate values of energy as expected.

Focusing now on the radiation emission curves for the different temperatures, we
observe that, for a low value of $\Lambda$, i.e. $\Lambda=0.1$, and in accordance to
the behaviour depicted in Fig. \ref{Temp_plot}, the EER for the normalised temperature
$T_{BH}$ is clearly the dominant one [see, Fig. \ref{EER_xi0}(a)]; the ones for the bare
$T_0$ and effective $T_{effEIW}$ temperatures follow behind, while for the remaining
effective ones -- that all take a very small value in the low-$\Lambda$ regime -- the EERs
are severely suppressed. For a value of $\Lambda$ close to its critical one, namely for
$\Lambda=0.8$, the only significant EERs are now the ones for $T_{BH}$ and $T_{eff-}$,
as one may see in Fig. \ref{EER_xi0}(b): all the other temperatures adopt a much smaller
value near the critical limit, and the corresponding spectra are thus suppressed.

\begin{figure}[t!]
  \begin{center}
\mbox{\includegraphics[width = 0.49 \textwidth] {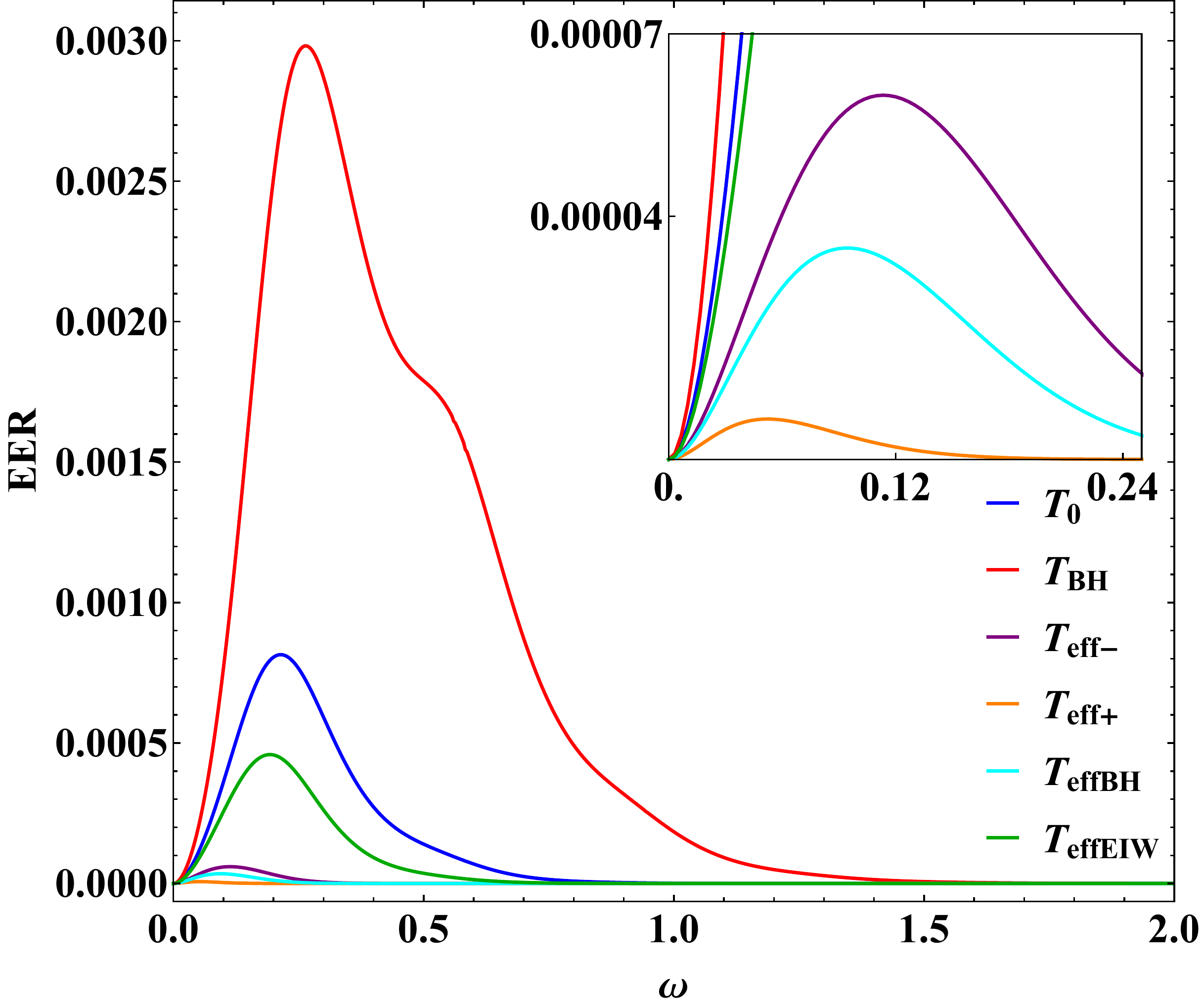}}
\hspace*{-0.1cm} {\includegraphics[width = 0.47 \textwidth]
{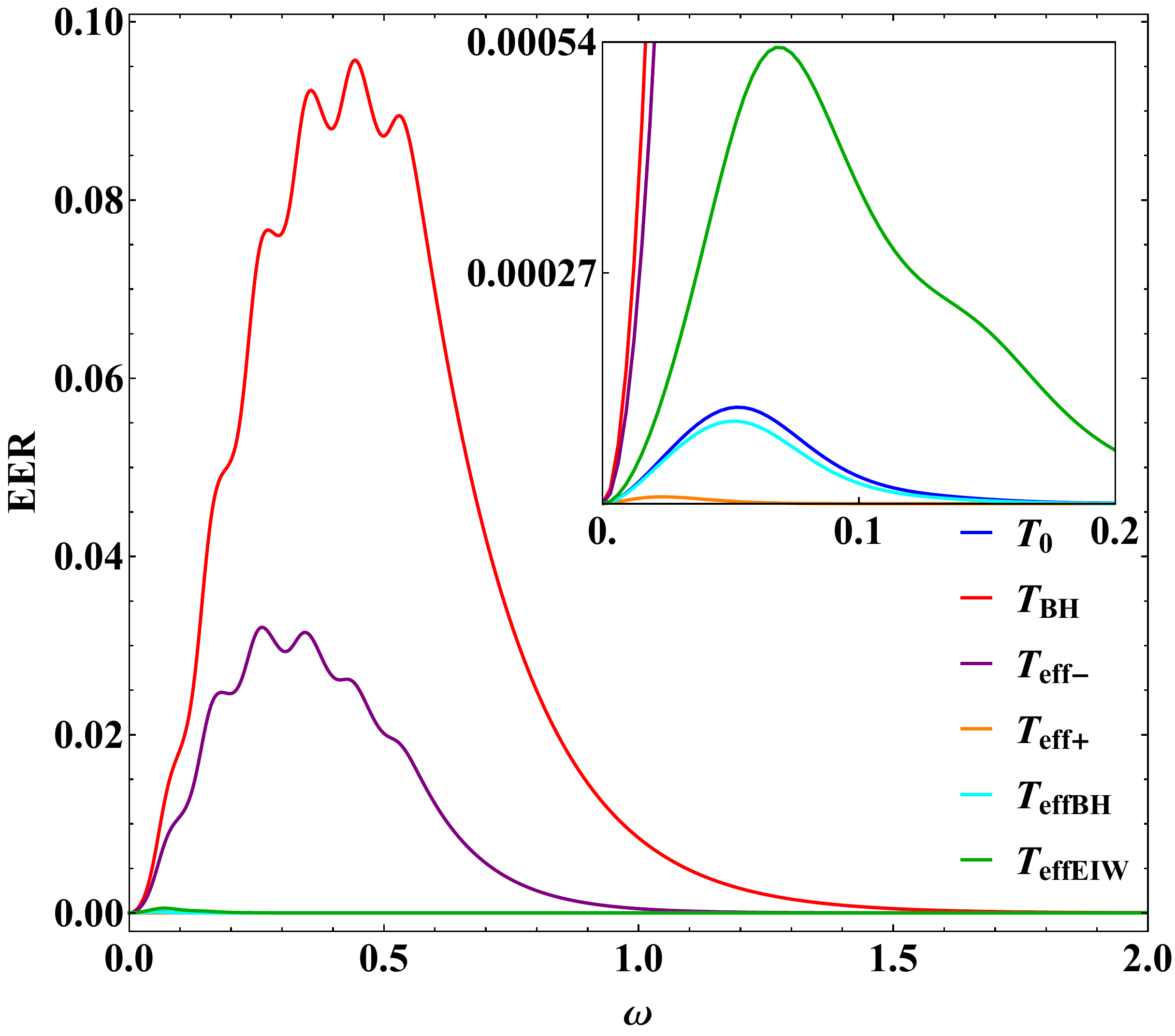}}
    \caption{Energy emission rates for non-minimally-coupled scalar fields with $\xi=1/6$
    from a Schwarzschild-de Sitter black hole for:  \textbf{(a)} $\Lambda=0.1$ (in units
    of $r_h^{-2}$), and  $T=T_{BH}$, $T_0$, $T_{effEIW}$, $T_{eff-}$, $T_{effBH}$,
    $T_{eff+}$ (from top to bottom), and \textbf{(d)} $\Lambda=0.8$ and $T=T_{BH},
    T_{eff-}$ (from top to bottom, again).}
   \label{EER_xi1/6}
  \end{center}
\end{figure}
\begin{figure}[b!]
  \begin{center}
\mbox{\includegraphics[width = 0.49 \textwidth] {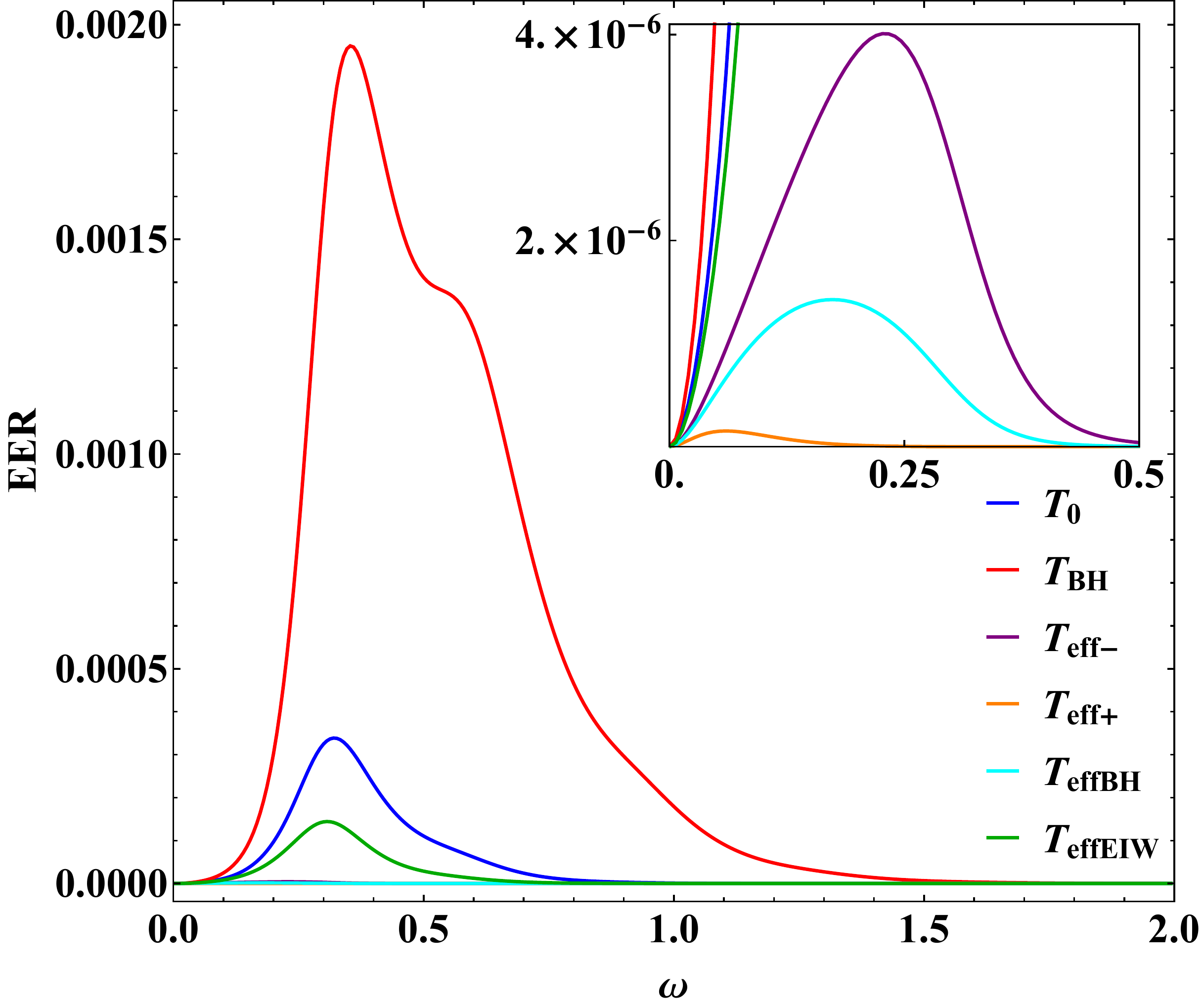}}
\hspace*{-0.1cm} {\includegraphics[width = 0.47 \textwidth]
{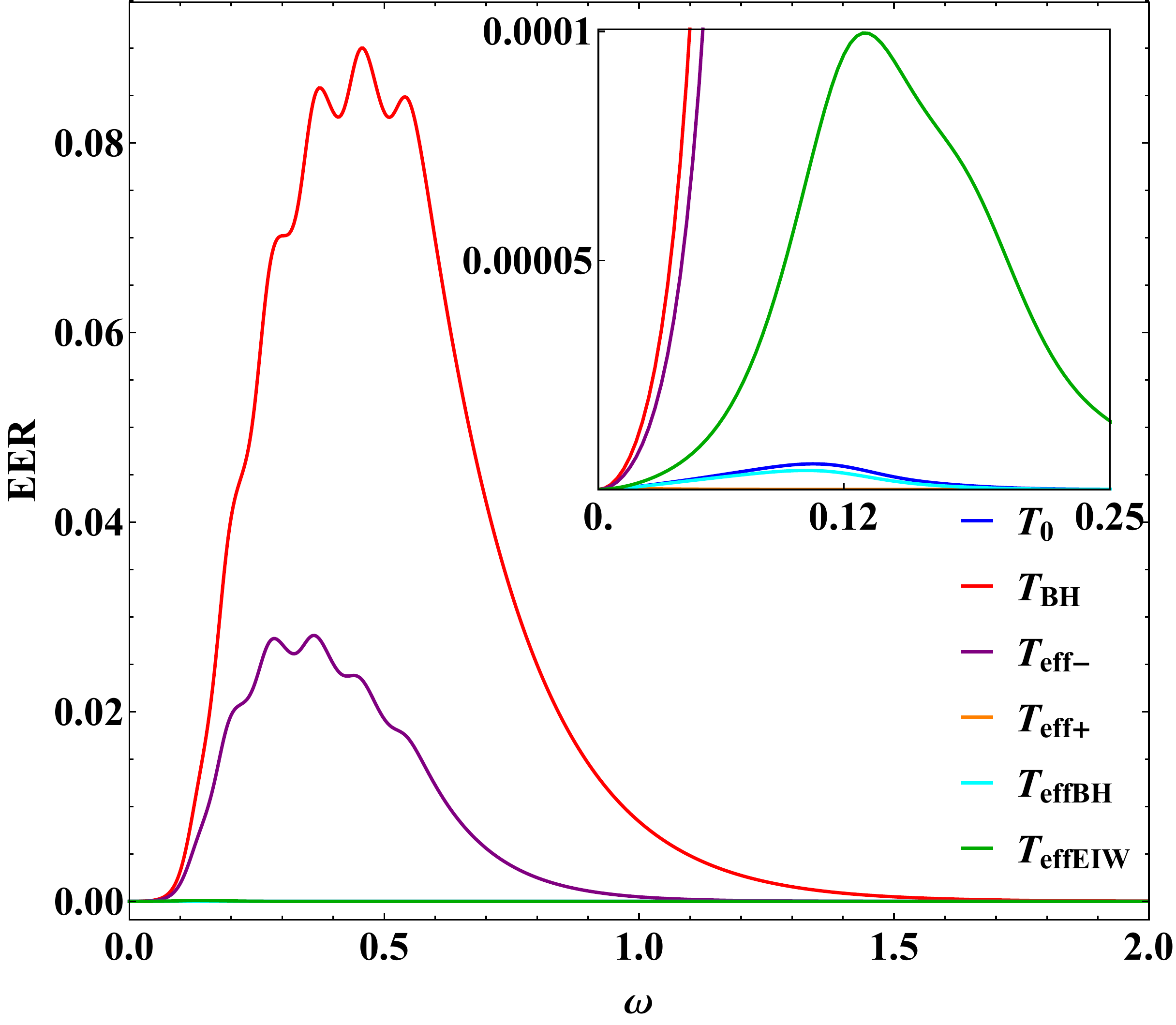}}
    \caption{Energy emission rates for non-minimally-coupled scalar fields with $\xi=1/2$
    from a Schwarzschild-de Sitter black hole for:  \textbf{(a)} $\Lambda=0.1$ (in units
    of $r_h^{-2}$), and  $T=T_{BH}$, $T_0$, $T_{effEIW}$, $T_{eff-}$, $T_{effBH}$,
    $T_{eff+}$ (from top to bottom), and \textbf{(d)} $\Lambda=0.8$ and $T=T_{BH},
    T_{eff-}$ (from top to bottom, again).}
   \label{EER_xi05}
  \end{center}
\end{figure}

Let us address now the case of a non-minimally coupled scalar field by assigning
a non-vanishing value to the coupling constant $\xi$. As soon as we do that, the effective
mass term re-appears and the non-zero low-energy asymptotic limit of the EERs
disappears. This is obvious from the plots in Fig. \ref{EER_xi1/6}, where the 
emission curves -- drawn for the indicative value of $\xi=1/6$ -- have now assumed
their traditional shape. The non-zero value of the coupling constant $\xi$ causes
a suppression in the energy emission rates over the whole energy regime: the 
emission of very low-energy particles has been severely suppressed, due to the
disappearance of the low-energy asymptotic limit, but also the peaks of the curves
are now lower. The same behaviour is observed if one increases further the value
of the coupling constant. In Fig. \ref{EER_xi05}, we depict the EERs for the
value $\xi=1/2$: all emission curves are further suppressed and the same pattern
continues for even higher values of $\xi$. What one could note is that the suppression
with $\xi$ is much stronger when the cosmological constant takes a small value,
while it becomes milder when $\Lambda$ approaches its critical limit - we will
return to this observation in the next section where the total emissivity of the
black hole is computed. 

The turning on of the non-minimal coupling constant $\xi$ is also in a position to
change the relative values of the EERs for the different temperatures, as is clear
from both Figs. \ref{EER_xi1/6} and \ref{EER_xi05}. However, the sequence and
general behaviour of the emission curves remains the same: for
small cosmological constant ($\Lambda=0.1$), it is the group of temperatures
($T_{BH}$, $T_0$, $T_{effEIW}$) -- i.e. the ones that adopt a non-vanishing value in
the limit $\Lambda \rightarrow 0$ -- that lead to significant EERs; close to the
critical limit ($\Lambda=0.8$), it is instead only $T_{BH}$ and $T_{eff-}$ that support
a non-vanishing energy emission rate for the black hole. Another common feature
is once again the dominance of the normalised temperature $T_{BH}$ over the whole
energy- and the whole-$\Lambda$ regime.


\section{Total Emissivities}
\label{sec:emiss}

In this section, we will calculate the total emissivity of the SdS black-hole, i.e. the
total energy emitted by the black hole per unit time over the whole energy regime,
in the form of scalar fields. We will see that this important quantity depends strongly
on the selected temperature with the results differing at times by orders of magnitude.  
Apart from choosing a different temperature each time, the total emissivities will
be computed for three values of the non-minimal coupling constant, i.e. for
$\xi=0,1/6,1/2$, and for four values of the cosmological constant, namely 
$\Lambda=(0.1,0.3,0.5,0.8)\,r_h^{-2}$. In this respect, our current analysis completes
and extends the analysis of \cite{Crispino} where only small values of $\Lambda$
were considered. Here, the values of $\Lambda$ span the whole allowed regime
up to its critical limit. This latter regime comprises in fact an important theoretical
limit, that is usually ignored in the calculation of Hawking radiation. That was due
to the fact that the traditional `bare' temperature $T_0$ vanishes at the critical limit
and the emission of Hawking radiation stops. However, as we showed, the normalised
temperature $T_{BH}$ as well as a number of the recently proposed effective temperatures
assume there a non-vanishing value. This leads not only to emission of Hawking radiation
but also to the maximization of the total emissivities for the SdS black hole in that limit.

The total emissivities for the SdS black hole, for the aforementioned values of $\xi$ and
$\Lambda$, are presented in Tables 1 through 3. Let us first examine how the increase in
the value of $\Lambda$ affects our results in the minimal coupling case ($\xi=0$). We
observe that, in accordance to the results of the previous sections, the total emissivity for 
the bare temperature $T_{0}$ decreases, and drops to only 14\% of its original value as
$\Lambda$ increases from 0.1 to 0.8. The total emissivity for the effective temperature
$T_{effEIW}$ also decreases, but now the decrease is milder, of the order of 73\%.
When the normalised temperature $T_{BH}$ is used, the total emissivity of the black
hole is actually enhanced, by a factor of 30 when $\Lambda$ increases from 0.1 to 0.8.
An increase is also noted for the total emissivity for $T_{eff-}$ but now the enhancement
factor is of the order of 300. Finally, the remaining two effective temperatures $T_{eff+}$
and $T_{effBH}$ see their emissivities first to increase and then to decrease as $\Lambda$
increases.

\begin{table}[t!]
\caption{Total emissivity for $\xi=0$} 
\centering 
\begin{tabular}{|c || c| c| c| c|} 
\hline\ 
$  \Lambda \rightarrow$ & 0.1 & 0.3 & 0.5 & 0.8  \\ [0.5ex] 
\hline 
\hline $T_0$& 0.000444 & 0.000487 & 0.000335 & 0.000065 \\ 
\hline$T_{BH}$& 0.001871 & 0.005432 & 0.011837 & 0.054554 \\
\hline$T_{eff-}$& 0.000058 & 0.000636 & 0.002106 & 0.015937 \\ 
\hline$T_{eff+}$& 0.000013 & 0.000047 & 0.000050 & 0.000014 \\ 
\hline$T_{effBH}$& 0.000040 & 0.000200 & 0.000225 & 0.000060 \\
\hline$T_{effEIW}$& 0.000266 & 0.000267 & 0.000222 & 0.000196 \\ 
\hline 
\end{tabular}
\label{tablexi0} 
\end{table}

\begin{table}[t!]
\caption{Total emissivity for $\xi=1/6$} 
\centering 
\begin{tabular}{|c || c| c| c| c|} 
\hline\ 
$  \Lambda \rightarrow$ & 0.1 & 0.3 & 0.5 & 0.8  \\ [0.5ex] 
\hline 
\hline $T_0$& 0.000228 & 0.000124 & 0.000057 & 7.5796 $10^{(-6)}$ \\ 
\hline$T_{BH}$& 0.001358 & 0.003647& 0.008889 & 0.050743 \\
\hline$T_{eff-}$& 9.8980 $10^{(-6)}$ & 0.000191 & 0.001040 & 0.013964 \\ 
\hline$T_{eff+}$& 0.5696 $10^{(-6)}$ & 1.3972 $10^{(-6)}$& 1.2237 $10^{(-6)}$ & 0.2977 $10^{(-6)}$ \\ 
\hline$T_{effBH}$& 5.0160 $10^{(-6)}$ & 0.000026 & 0.000027 & 6.3134 $10^{(-6)}$ \\
\hline$T_{effEIW}$& 0.000112 & 0.000044 & 0.000026 & 0.000054 \\ 
\hline 
\end{tabular}
\label{tablexi1/6} 
\end{table}

\begin{table}[t!]
\caption{Total emissivity for $\xi=1/2$} 
\centering 
\begin{tabular}{|c || c| c| c| c|} 
\hline\ 
$  \Lambda \rightarrow$ & 0.1 & 0.3 & 0.5 & 0.8  \\ [0.5ex] 
\hline 
\hline $T_0$& 0.000087 & 0.000021 & 6.1253 $10^{(-6)}$ & 0.5443 $10^{(-6)}$ \\ 
\hline$T_{BH}$& 0.000837 & 0.002126 & 0.006062 & 0.045459 \\
\hline$T_{eff-}$& 0.8973 $10^{(-6)}$ & 0.000040 & 0.000433 & 0.011571 \\ 
\hline$T_{eff+}$ & 0.0164 $10^{(-6)}$ & 0.0316 $10^{(-6)}$ & 0.0251 $10^{(-6)}$ & 0.0057 $10^{(-6)}$  \\ 
\hline$T_{effBH}$& 0.3206 $10^{(-6)}$ & 1.7945 $10^{(-6)}$ & 1.8766 $10^{(-6)}$ & 0.4070 $10^{(-6)}$  \\
\hline$T_{effEIW}$& 0.000033 &  4.1488 $10^{(-6)}$ & 1.8064 $10^{(-6)}$ & 0.000012 \\ 
\hline 
\end{tabular}
\label{tablexi1/2} 
\end{table}

As the non-minimal coupling constant increases from $\xi=0$ to $\xi=1/6$ and then
to $\xi=1/2$, all total emissivities decrease since the appearance of the effective mass
term suppresses the EERs at all energy regimes. What is more significant is that the
value of $\xi$ strongly affects the suppression or enhancement factors for the total
emissivities -- computed for the different temperatures -- as $\Lambda$ varies.
For example,
the total emissivity for the bare $T_0$ drops to 3\%, for $\xi=1/6$, and to only
0.6\%, for $\xi=1/2$, of its original value as $\Lambda$ goes from 0.1 to 0.8.
Similarly, the total emissivity for the effective $T_{effEIW}$ drops to 48\% and 33\%,
respectively. On the other hand, the total emissivity for
the dominant $T_{BH}$ is now enhanced by a factor of 37, for $\xi=1/6$, and by a
factor of 54, for $\xi=1/2$. The total emissivity for $T_{eff-}$ is the one that is
mostly affected: it increases by a factor of 1400, for $\xi=1/6$, and by a factor
of 12000, for $\xi=1/2$, as $\Lambda$ goes from 0.1 to 0.8. We expect the same
pattern to continue as $\xi$ increases further.

\section{Conclusions}
\label{sec:Conclusions}

In this work, we have considered the Schwarzschild-de Sitter black hole
and performed a study of the Hawking radiation spectra, emitted in the
form of scalar fields either minimally or non-minimally coupled to gravity.
The novel feature of our analysis is the use of six different temperatures
for the SdS background, as the question of the proper temperature for such a
spacetime is still debated in the literature. We have thus considered the
{\it bare} temperature, defined in terms of the black-hole surface gravity, the
{\it normalised} temperature, that takes into account the absence of an
asymptotically-flat limit, and four {\it effective} temperatures defined in
terms of both the black-hole and cosmological horizon temperatures. 

We first studied the profiles of the above temperatures as a function of the
cosmological constant $\Lambda$, from a zero value up to its maximum, critical
limit. We have found that the temperatures are split in two groups depending
on their behaviour in these two asymptotic $\Lambda$-regimes. In the limit
of zero cosmological constant, the aforementioned temperatures either reduce
to the temperature of the Schwarzschild black hole or vanish; near the critical
limit, they either assume a non-vanishing asymptotic value or reduce again to
zero.

Their different profiles inevitably affect the form of the energy emission 
rates for Hawking radiation. For small values of $\Lambda$, it is only the
bare $T_0$, the normalised $T_{BH}$ and the effective temperature $T_{effEIW}$
that lead to significant radiation from the SdS black hole. In the opposite
critical limit, it is only the spectra for $T_{BH}$ and $T_{eff-}$ that survive,
with the one for the normalised $T_{BH}$ being the dominant one over the whole
energy regime. The computation of the total emissivities confirm the above
behaviour in a quantitative way, and reveal that the value of the non-minimal
coupling constant $\xi$ determines the relative values of the EERs for the
different temperatures as well as the enhancement or suppression factors
for each one of them as the cosmological constant increases.

{\bf Acknowledgement} T.P. would like to thank the Alexander S. Onassis Public
Benefit Foundation for financial support.



\bibliographystyle{utphys}

\end{document}